\documentclass{article}

\usepackage[preprint]{neurips_2024}

\usepackage[utf8]{inputenc} 
\usepackage[T1]{fontenc}    
\usepackage{hyperref}       
\usepackage{url}            
\usepackage{booktabs}       
\usepackage{amsfonts}       
\usepackage{nicefrac}       
\usepackage{microtype}      
\usepackage{xcolor}         
\usepackage{tabularx}
\usepackage{graphics}
\usepackage{graphicx}
\usepackage{geometry}
\usepackage{adjustbox}
\usepackage{subcaption}
\usepackage{float}
\usepackage{multirow}
\usepackage{amsmath}
{

      \newtheorem{lemma}{Lemma}
      \newtheorem{definition}{Definition}
}

\usepackage{algorithm}
\usepackage{algpseudocode}
\algrenewcomment[1]{\hfill\textcolor{blue}{/* #1 */}}

\definecolor{darkgreen}{RGB}{0, 140, 0}
\definecolor{darkred}{RGB}{220, 20, 60}

\newcommand{\method}{\textsl{DREW}}
\newcommand{\methodexpand}{\underline{D}ata \underline{R}etrieval with \underline{E}rror-corrected codes and \underline{W}atermarking}

\title{\method: Towards Robust Data Provenance by Leveraging Error-Controlled Watermarking}

\author{
Mehrdad Saberi$^1$, Vinu Sankar Sadasivan$^{1}$, Arman Zarei$^{1}$, Hessam Mahdavifar$^2$, Soheil Feizi$^{1}$\\
$^1$ University of Maryland \quad $^2$ Northeastern University\\
}

\begin{document}

\maketitle

\begin{abstract}

Identifying the origin of data is crucial for data provenance, with applications including data ownership protection, media forensics, and detecting AI-generated content. A standard approach involves embedding-based retrieval techniques that match query data with entries in a reference dataset. However, this method is not robust against benign and malicious edits. To address this, we propose \methodexpand{} (\method{}). \method{} randomly clusters the reference dataset, injects unique error-controlled watermark keys into each cluster, and uses these keys at query time to identify the appropriate cluster for a given sample. After locating the relevant cluster, embedding vector similarity retrieval is performed within the cluster to find the most accurate matches. The integration of error control codes (ECC) ensures reliable cluster assignments, enabling the method to perform retrieval on the entire dataset in case the ECC algorithm cannot detect the correct cluster with high confidence. This makes \method{} maintain baseline performance, while also providing opportunities for performance improvements due to the increased likelihood of correctly matching queries to their origin when performing retrieval on a smaller subset of the dataset. Depending on the watermark technique used, \method{} can provide substantial improvements in retrieval accuracy (up to 40\% for some datasets and modification types) across multiple datasets and state-of-the-art embedding models (e.g., DinoV2, CLIP), making our method a promising solution for secure and reliable source identification. The code is available at 
\textcolor{magenta}{\href{https://github.com/mehrdadsaberi/DREW}{https://github.com/mehrdadsaberi/DREW}}.

\end{abstract}

\section{Introduction}

In the era of big data, the ability to accurately trace the provenance of data is critical for ensuring the integrity, reliability, and accountability of information \citep{provenance1, provenance2}. Data provenance, the documentation of the origins and the life-cycle of data, is essential for various applications, including the detection of AI-generated content and deepfakes, legal compliance, data ownership protection, and copyright protection. Some current approaches to data provenance often rely on metadata or external logs \citep{c2pa}, which can be easily tampered with or lost. To overcome these limitations, this paper introduces a novel approach that leverages error-controlled watermarking and retrieval techniques to address source identification in data provenance. 

Watermarking is a powerful technique for embedding information directly within data, enabling tracking of data origins \citep{watermark1, watermark2}. However, recent studies \citep{saberi2024robustness, zhang2023watermarks, rabbani2024waves, sadasivan2024aigenerated} have revealed significant limitations in watermark robustness against data augmentations. These studies show that watermarks often fail to preserve information when subjected to modifications such as cropping, blurring, diffusion purification \citep{saberi2024robustness, nie2022diffusion}, and adversarial attacks \citep{chakraborty2018adversarial}.

To mitigate information loss, some approaches \citep{stegastamp, trustmark} use error correction codes \citep{hamming1950error} like BCH codes \citep{bch} to recover lost information during transformations. However, the volume of information that can be reliably embedded as a watermark is limited (e.g., $100$-$200$ bits), restricting the inclusion of sufficient redundant bits for effective error correction. Consequently, existing watermarking methods are not suitable for source identification in large-scale scenarios. As a toy example, to handle billion-scale datasets, more than $2^30$ unique watermark keys are required, and with $100$-bit watermarks injected into the data, ECC algorithms cannot provide a highly robust coding.

Embedding-based retrieval techniques \citep{9933854} use data embedding similarities for provenance. These methods map data into a low-dimensional space where similar items are closer, allowing query embeddings to be compared against a database to find matches. These techniques have been effectively used in image retrieval \citep{clip, dinov2} and natural language processing \citep{bert, t5, gpt3} to infer data origins.

While embedding-based retrieval is robust against certain transformations, challenges arise as the reference database grows (see Figure~\ref{fig:k_analysis}) or when encountering challenging data alterations. These factors can impact retrieval accuracy, necessitating advancements in embedding techniques for enhanced performance and reliability. Additionally, embedding models, like other deep learning models, often struggle with generalization to out-of-distribution or under-represented data.

\begin{figure}
    \vspace{-0.7cm}
    \includegraphics[width=1\linewidth]{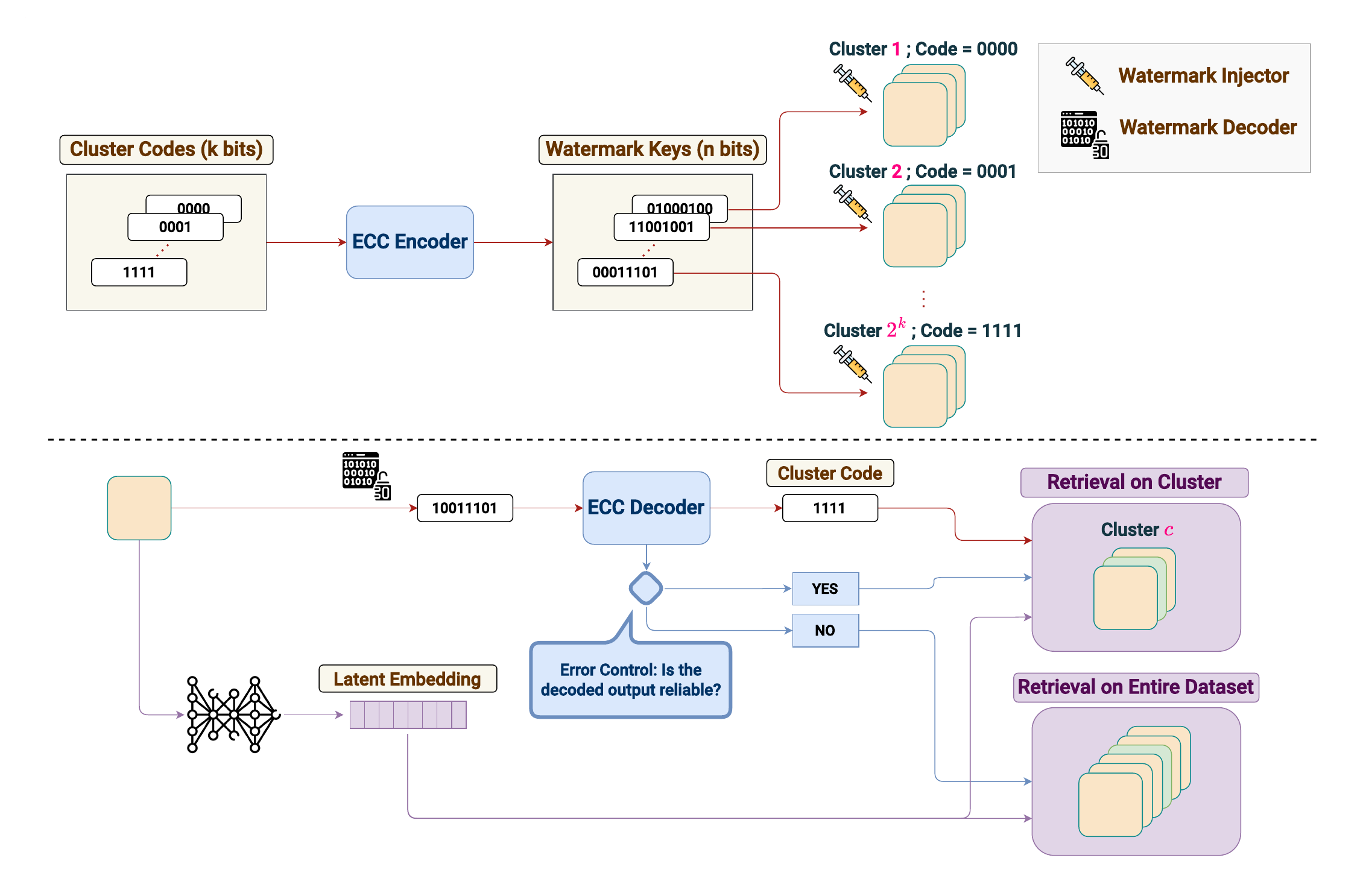}
    \caption{Overview of \method{}. \textbf{(Top)} During the pre-processing phase, $2^k$ clusters are created, each associated with a unique watermark key produced by the ECC encoder module. Instances from the dataset are randomly allocated to these clusters, and the corresponding watermark keys are injected into them. \textbf{(Bottom)} Upon receiving a query sample, the watermark decoder extracts the injected key, and the embedding model computes an embedding vector for the query. The ECC decoder processes this extracted key to identify the cluster code associated with the query. If the ECC reliability module confirms the reliability of the decoded cluster code, an embedding-based retrieval is conducted within the corresponding cluster. If the reliability is not ensured, the retrieval is performed across the entire dataset.
    }
    \label{fig:main_fig}
    \vspace{-0.5cm}
\end{figure}

\begin{figure}
    \centering
    \includegraphics[width=1.0\textwidth]{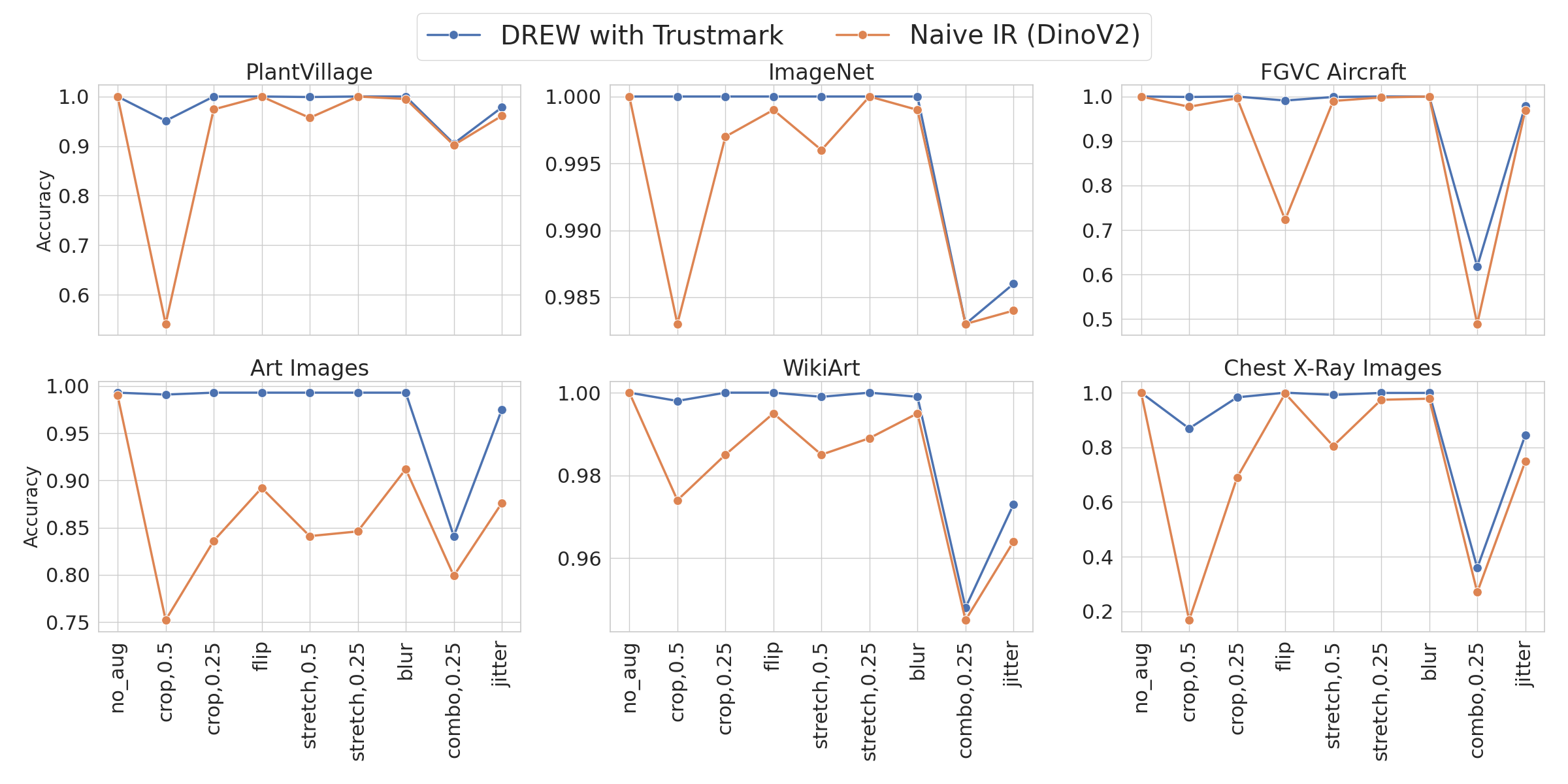} \\
    \caption{Source identification accuracy of \method{} (with Trustmark used as the watermarking technique) vs naive embedding-based retrieval using DinoV2 embeddings, against different types of augmentations.}
    \label{fig:acc_dinov2}
    \vspace{-0.3cm}
\end{figure}

This paper introduces a novel approach, \method{} (\methodexpand{}), that integrates watermarks, error control codes, and embedding-based retrieval to address the challenge of source identification. The proposed framework effectively mitigates the limitations of watermark capacity (i.e., the constraints on the number of bits that can be stored within content), and the scalability issues inherent in data retrieval methods (i.e., the decreasing accuracy of retrieval in larger reference databases). By leveraging the complementary strengths of watermarking and retrieval techniques, \method{} presents a more robust alternative for source identification.

In \method{}, data from the reference dataset is randomly clustered, with each cluster assigned a unique $k$-bit binary cluster code. To enhance robustness, we use an ECC encoder to add redundancy to these codes, resulting in new $n$-bit keys for each cluster. These keys are then injected into the data samples from the corresponding cluster using a watermark injector. The watermarked samples, which can be any type of media and content such as outputs of generative models, images posted on social media platforms, or legal documents, are subsequently published to general users. If a watermarked sample from our reference dataset is modified, we can reverse the process to identify the corresponding cluster and perform embedding-based data retrieval within that cluster. Additionally, the ECC module provides a reliability flag, enabling us to perform retrieval on the entire reference dataset if the retrieved cluster code is deemed unreliable due to excessive data modification.

\method{} offers several key advantages:
\begin{itemize}
    \item \textbf{Robust Watermark Utilization:} By using $k$-bit cluster codes and encoding them using ECC to create $n$-bit watermark keys, we can adjust $k$ as a hyperparameter to control the ECC decoder's robustness. This flexibility is crucial because using unique $n$-bit watermark keys for each data instance in a large-scale dataset would make robust decoding impractical.

    \item \textbf{Improved Retrieval Accuracy:} Conducting data retrieval on smaller clusters enhances accuracy, addressing challenges associated with purely embedding-based retrieval.

    \item \textbf{Reliability of ECC Algorithms:} Our method might underperform compared to purely embedding-based retrieval only if the ECC algorithm returns an incorrect cluster code and falsely marks it as reliable. However, modern ECC algorithms have near-optimal coding efficiency with very low false positive rates, ensuring that our method maintains performance on par with the baseline.
\end{itemize}

The following sections provide a detailed formal description of our method, along with theoretical insights and empirical analysis on the impact of the different modules utilized in our framework. Furthermore, we conduct comprehensive experiments to demonstrate the effectiveness of our approach compared to the baseline of performing embedding-based retrieval on the entire dataset. While our framework is applicable to various data types, we focus our experiments on the image domain.

As shown in Figure~\ref{fig:acc_dinov2}, in practice, \method{} provides significant robustness, especially on more challenging datasets where naive embedding-based retrieval methods struggle. In Section~\ref{sec:experiments}, we show that while \method{} shows substantial performance gains against some augmentations, it does not cause performance degradation against any modification types, when compared to the baseline.

Below, we list the main contributions of this paper:
\begin{itemize}
\item We propose \method{}, a novel framework for source identification, by combining error correction codes (ECC), watermarking techniques, and embedding-based data retrieval. Our framework addresses the limitations of using watermarks or data retrieval independently, providing a robust and scalable solution by combining their strengths.
\item We provide theoretical insights into the effectiveness of our framework, followed by empirical results supporting our claims.
\item We conduct extensive experiments on multiple image datasets, showcasing the improved robustness of our method compared to the baseline of embedding-based data retrieval using state-of-the-art embedding models (e.g., DinoV2 \citep{dinov2}, CLIP/ViT-L-14 \citep{clip}).
\end{itemize}





\section{Background}

\textbf{Watermarking Techniques.} Watermarking involves embedding information within content so that it can be reliably decoded later, even after the content has been modified. This technique has been applied to various types of media, including images \citep{WAN2022226}, text \citep{kirchenbauer2024watermark}, and audio \citep{HUA2016222}. Recent advancements in watermarking methods utilize deep learning models for encoding and decoding, leading to improved robustness in the face of several types of content modifications \citep{trustmark, stegastamp, treering, kirchenbauer2024watermark}. However, numerous studies have highlighted the limitations and unreliability of these watermarks when subjected to certain data augmentations and transformations \citep{saberi2024robustness, sadasivan2024aigenerated, zhang2023watermarks, rabbani2024waves}. As a result, relying solely on watermarking as a solution can be challenging in many scenarios.

\textbf{Error Control Codes.} Error control codes (ECC) \citep{hamming1950error} are fundamental in ensuring the reliable transmission and storage of data across noisy channels. These codes detect and correct errors without needing re-transmission, enhancing data integrity and communication efficiency. Notable ECCs include Reed-Solomon codes \citep{reedsolomon}, known for robust error correction in digital communications and storage like CDs and DVDs. LDPC codes \citep{ldpc} have transformed ECC with near-capacity performance, crucial in modern systems like 5G and satellite communications. BCH codes \citep{bch}, introduced in the 1950s, are versatile for error detection and correction in applications from QR codes to digital broadcasting. Polar codes \citep{polar_codes}, a recent advancement, achieve channel capacity with low-complexity decoding, essential for 5G technology. These advancements highlight ECC's evolution, driven by the need for high reliability and efficiency in complex communication networks.

\textbf{Data Retrieval.} Data retrieval techniques refer to the methods and algorithms used to efficiently search, access, and retrieve relevant information from large data repositories or databases \citep{9933854}. Recent advancements include embedding-based retrieval systems, which utilize deep learning models to transform data into feature vectors, enabling accurate and context-aware searches. Approximate Nearest Neighbors (ANN) \citep{10.1145/293347.293348, simhadri2022results} is a notable technique in this field, facilitating rapid similarity searches by approximating the nearest neighbor search in large datasets. Data retrieval techniques are especially useful in applications such as recommendation systems \citep{naumov2019deep}, retrieval augmented generation \citep{lewis2021retrievalaugmented}, near-duplicate detection \citep{6553136}, and anomaly detection \citep{cohen2021subimage}.

\textbf{Watermarking for Data Retrieval.} Some existing works have explored the use of watermarks in retrieval-based tasks in various ways. Certain studies \citep{roy2005unified, roy2004watermarking} propose manually shifting data (similar to watermarking) to enhance the mutual separation of samples in feature space, thereby improving the robustness of feature-based data retrieval. However, these methods do not leverage recent advancements in watermarking or efficient retrieval techniques, rendering them impractical compared to contemporary approaches. Additionally, other works \citep{anju2022faster, lafta2022secure} have utilized watermarks in image retrieval systems for user identification and to prevent illegal data distribution. Nonetheless, these approaches fail to address the low robustness of watermarks when used alone and without error correction modules, making them unreliable in practical applications.


\section{\method{}: Data Retrieval with Error-Corrected Codes and Watermarking}


In the following sections, we formally define the problem and present our approach. We note that the framework in this section is not dependent on the type of data and can be potentially applied to data types such as images, text, audio, etc. However, our experiments in Section~\ref{sec:experiments} are performed in the image domain.


\begin{algorithm}
\caption{Preprocessing for \method{}}
\label{alg:method-pre}
\begin{algorithmic}[1]
\Require $X$: dataset, $k$: bit length of cluster codes, $\mathcal{C}_E$: error control encoder, $\mathcal{W}_I$: watermark injector,  $\phi$: embedding model
\State Randomly partition $X$ into $2^k$ clusters $X_0, ..., X_{2^k-1}$
\State $\forall_{i=0}^{2^k-1} \ c[i] \ \gets \ $ $k$-bit binary representation of number $i$ \Comment{Set cluster codes}
\For{$i$ : $0 \rightarrow 2^k-1$}
\State $w \ \gets \ \mathcal{C}_E$($c[i]$)
\For{$x$ in $X_i$}
\State $x' \ \gets \ \mathcal{W}_I(x, w)$
\State replace $x$ with $x'$ in $X$
\State Store $\phi(x)$ for query phase

\EndFor
\EndFor
\State \textbf{return} $X=\bigcup\limits_{i=0}^{2^k-1} X_i, \ \phi(X)$  \Comment{return watermarked clusters and data embeddings}
\end{algorithmic}
\end{algorithm}

\subsection{Problem Definition}
Consider a dataset $X=\{x_0, x_1, ..., x_{N-1}\}$ containing images or other types of data. New instances, denoted by $z \in \mathbb{R}^D$, may be added to this dataset. The primary objective is to detect whether $z$ is either an exact duplicate or a modified version of any data currently existing within $X$. The allowed modifications are a set of alterations that do not change the semantics of the data, and the modified data is considered a copy of the original data based on human judgment. In the case where a matching data sample is found, the method is required to identify and report this instance. Conversely, the method should also report if no such match exists. For some applications, we might want to proceed to add $z$ to the dataset after finding no matching sample for it. Typically, this problem is addressed through the application of data retrieval techniques.

\subsection{Preliminaries}
\label{sec:prelims}

The data modification function (i.e., attack function) is shown as $\mathcal{A}: \mathbb{R}^D \rightarrow \mathbb{R}^D$, which can be performed on $x_i \in X$ to create a query $z=\mathcal{A}(x_i)$.
A method $ \mathcal{M} : \mathbb{R}^D \rightarrow \mathbb{R}^D$ that addresses the problem, receives a query sample $z$, and outputs a matching data instance from $X$. Note that for simplicity of the analysis in this section, we assume that methods always output a match for all the queries. Further analysis of scenarios where queries do not necessarily have a corresponding match in the dataset is provided in Appendix~\ref{appendix:out-of-dataset}.


\textbf{Embedding Model.} We utilize a data embedding model $\phi : \mathbb{R}^D \rightarrow \mathbb{R}^d$, to create a normalized low-dimensional embedding of the data. This embedding is used to measure the similarity between two data samples $z_1$ and $z_2$, by calculating the dot product between their embeddings (i.e., $\phi(z_1)^\top \phi(z_2)$).

\textbf{Watermarking Module.} The watermarking module consists of two models. The watermark injector $\mathcal{W}_I: \mathbb{R}^D \times \{0,1\}^n \rightarrow \mathbb{R}^D$ receives a data sample and a $n$-bit binary key, and outputs a data sample which contains the key embedded inside of it. The watermark decoder $\mathcal{W}_D: \mathbb{R}^D \rightarrow \{0,1\}^n$ receives a data sample and retrieves the watermark key that is embedded in it.

\textbf{Error Control Module.} The error control module \citep{shannon_theorem, hamming1950error} consists of multiple modules. There is an encoder $\mathcal{C}_E: \{0,1\}^k \rightarrow \{0,1\}^n$ that receives a $k$-bit binary code and outputs an $n$-bit binary code. This $n$-bit binary code can be passed through a noisy channel (i.e., a channel that can alter the bits), which for our case, we consider a channel that performs randomized flip operation on the bits (i.e., changes some $0$ bits to $1$, and vice versa). 

The decoder module $\mathcal{C}_D: \{0,1\}^n \rightarrow \{0,1\}^k$ receives the $n$-bit binary code, after it has been passed through the channel, and outputs the initial $k$-bit code from which the $n$-bit code was created.

The last module is reliability check module $\mathcal{C}_R: \{0,1\}^n \rightarrow \{0,1\}$ that receives an $n$-bit binary code, and outputs a reliability flag that determines whether $\mathcal{C}_D$ can confidently decode this input, or if the noise that was applied to the $n$-bit code in the channel was more than the decoder's capacity of correction (a value of $1$ represents a reliable output).

\subsection{Methodology: \method}
\label{sec:method}





\begin{algorithm}
\caption{Querying with \method{}}
\label{alg:method-query}
\begin{algorithmic}[1]
\Require $z$: query sample, $X$: watermarked dataset, $\mathcal{W}_D$: watermark decoder, $\mathcal{C}_D$: error control decoder, $\mathcal{C}_R$: error control reliability module, $\phi$: embedding model, $\tau_r$: retrieval threshold
\State $w \ \gets \ \mathcal{W}_D(z)$
\State $c \gets \ \mathcal{C}_D (w)$
\State $r \gets \ \mathcal{C}_R (w)$
\If{$r$} \Comment{if the output of error control decoder is reliable}
\State $S(z) \gets X_c$
\Else
\State $S(z) \gets X$
\EndIf
\State $x^* \gets \mathop{\arg\max}_{x \in S(z)} \phi(x)^\top \phi(z)$
\If{$\phi(x^*)^\top \phi(z) < \tau_r$} \Comment{if similarity is low, no match has been found}
\State \textbf{return} -1
\EndIf
\State \textbf{return} $x^*$ \Comment{if similarity is higher than threshold, return the match}
\end{algorithmic}
\end{algorithm}

\textbf{Preprocess.} We randomly partition the data from $X$ into $2^k$ clusters and assign a unique binary code with length $k$ to each cluster (i.e., cluster codes). Next, we use the encoder model $\mathcal{C}_E$ of an error control module to create an $n$-bit binary key for each cluster (i.e., cluster watermark keys). These $n$-bit binary keys are injected as watermarks into the data samples in each cluster (using the watermark injector model $\mathcal{W}_I$). The watermarked dataset replaces the original dataset (i.e., $X$ now contains watermarked data). Algorithm~\ref{alg:method-pre} shows the pseudocode for \method{}'s pre-processing.

\textbf{Query.} When a query sample $z$ (which could be a modified version of an existing data sample in $X$), we use the watermark decoding model $\mathcal{W}_D$, which gives us an $n$-bit binary watermark key as output (i.e., $w = \mathcal{W}_D(z)$). Next, we use the error control decoder $\mathcal{C}_D$ to retrieve the $k$-bit cluster code that corresponds to the cluster in which the non-altered version of $z$ belongs (i.e., $c = \mathcal{C}_D(w)$). If $\mathcal{C}_R$ flags the output of decoder as unreliable (i.e., $\mathcal{C}_R(w)=0$), we do not rely on the decoded cluster code and set our search space $S(z)$ to be equal to $X$ (i.e., the dataset which includes watermarked samples after the pre-processing phase).
Otherwise, if the decoded cluster code was reliable, we set the search space $S(z)$ to only include samples
from the cluster with the corresponding cluster code.

Next, we utilize the embedding model $\phi$ to find the data sample in $S(z)$ with the highest similarity to $z$. Since the embedding-based data retrieval is potentially being performed on a small subset of $X$, it is expected to observe improvements in both accuracy and speed, compared to performing naive data retrieval on all of $X$. The pseudocode for the query procedure of the method is shown in Algorithm~\ref{alg:method-query}.

\subsection{Performance Analysis}
\label{sec:performance_analysis}


We consider $\mathcal{M}_{wm}$ to be our proposed method \method{} (refer to Section~\ref{sec:method}), and $\mathcal{M}_\phi$ to be the baseline method of performing embedding similarity check on the entire dataset.

For the analysis in this section, we focus on queries that have a ground truth match in the dataset. We will compare the accuracy of detecting the correct match between $\mathcal{M}_{wm}$ and $\mathcal{M}_{\phi}$ for $z$, where $z=\mathcal{A}(x_i)$ and $x_i \in X$. We assume that the prediction of $\mathcal{C}_D$ for the cluster code is $c=\mathcal{C}_D(\mathcal{W}_D(z))$, and the reliability flag outputted by $\mathcal{C}_R$ is $r=\mathcal{C}_R(\mathcal{W}_D(z))$. Additionally, the correct cluster that includes $x_i$ is $c_{gt}$ (i.e., $x_i \in X_{c_{gt}}$).
Now, if $r=1$, the output of $\mathcal{M}_{wm}$ is $x^*=argmax_{x \in X_{c_{gt}}} \phi(x)^\top \phi(z)$.
The outputs of $\mathcal{M}_\phi$, or $\mathcal{M}_{wm}$ when $r=0$, are $\bar{x}^*=argmax_{x \in X} \phi(x)^\top \phi(z)$.

\begin{definition}
\label{def:ecc_error}
    We define $\epsilon_r$ to be the error of the error control decoder $\mathcal{C}_D$ in outputting the correct $k$-bit cluster code, in cases where $\mathcal{C}_R$ has an output of $1$. Formally,
    $$\epsilon_r = \mathbb{P}(c \neq c_{gt} | r=1).$$
    \label{def:ecc_fp}
\end{definition}
\vspace{-0.3cm}
Based on Definition~\ref{def:ecc_fp}, the false positive rate of $\mathcal{C}_R$ is $\epsilon_r$. In Figure~\ref{fig:epsilon_r}, we show that with the correct setting, $\epsilon_r$ has a small and insignificant value.

For a method $\mathcal{M}$, an attack $\mathcal{A}$, and a dataset $X$, we define the error in identifying the correct match as $E(\mathcal{M}, \mathcal{A}, X)$.
We can analyze the difference in error of $\mathcal{M}_{wm}$ and $\mathcal{M}_\phi$ as follows:
\begin{equation}
\label{eq:improvement}
\begin{split}
 E(\mathcal{M}_{\phi},\mathcal{A},X) & - E(\mathcal{M}_{wm},\mathcal{A},X) \\
 & = \mathbb{P}(x^*=x_i, \bar{x}^* \neq x_i, r=1, c=c_{gt}) - \mathbb{P}(c \neq c_{gt}, r=1) \\
& \geq \mathbb{P}(r=1, c=c_{gt} | x^*=x_i , \bar{x}^* \neq x_i) \mathbb{P}(x^*=x_i , \bar{x}^* \neq x_i) - \epsilon_r
\end{split}
\end{equation}
The term $\mathbb{P}(r=1, c=c_{gt} | x^=x_i , \bar{x}^ \neq x_i)$ reflects the robustness of the watermark and ECC in producing the correct output against an attack, compared to the robustness of the embeddings against the same attack. Meanwhile, $\mathbb{P}(x^=x_i , \bar{x}^ \neq x_i)$ indicates the improvement from reducing the dataset samples needed for embedding-based retrieval. In the following lemma, we measure $\mathbb{P}(x^=x_i , \bar{x}^ \neq x_i)$ under certain assumptions on embedding-based retrieval:
\begin{lemma}
    Assume that the top-1 accuracy of the embedding-based retrieval on dataset $X$ with size $N$, is $\alpha$ (i.e., $\mathbb{P}(\bar{x}^* = x_i)=\alpha$), and its top-$p$ accuracy is $\alpha_p$. Then,
    $$\mathbb{P}(x^*=x_i , \bar{x}^* \neq x_i) \geq \mathop{max}_{p \geq 2} [(\alpha_p-\alpha) (1-\frac{1}{2^k})^{p-1}].$$
    \label{lem:top-p}
\end{lemma}
Based on Lemma~\ref{lem:top-p}, our method archives the highest accuracy boost in cases where there is a significant gap between the top-1 and top-$p$ accuracy of the embedding-based retrieval for some value of $p$ that is not too large. While $\alpha_p - \alpha$ increases for larger values of $p$, the term $(1-\frac{1}{2^k})^{p-1}$ prevents $p$ from having large values, introducing a trade-off.

In Figure~\ref{fig:epsilon_r} and Figure~\ref{fig:k_analysis}, we provide empirical analysis on terms from Equation~\ref{eq:improvement}. According to Figure~\ref{fig:epsilon_r}, $\epsilon_r$ has an insubstantial value on practice, and does not limit the performance improvement of \method{}. Furthermore, Figure~\ref{fig:k_analysis} demonstrates that the term $\mathbb{P}(x^*=x_i , \bar{x}^* \neq x_i)$ has a noticeable value for suitable values of $k$. Therefore, we can conclude that the bottleneck of \method{} is the term $\mathbb{P}(r=1, c=c_{gt} | x^*=x_i , \bar{x}^* \neq x_i)$, which corresponds to the relative robustness of the watermarking technique used, compared to the robustness of the embeddings. This implies that by designing more robust watermarks, we can further increase the performance gap between \method{} and traditional embedding-based retrieval.

In the appendix, we provide further analysis on the robustness of the ECC module (Appendix~\ref{appendix:ecc}), the robustness of \method{} against adversarial attacks (Appendix~\ref{appendix:adversarial}), and the performance of \method{} in the presence of out-of-dataset queries (Appendix~\ref{appendix:out-of-dataset}).

\begin{table}
\vspace{-0.2cm}
\centering
\begin{adjustbox}{max width=\textwidth}
\begin{tabular}{llcccccccccc}
\toprule
Dataset & Method & No Aug & Rot 1.0 & Rot 0.5 & Rot 0.25 & Crop 0.5 & Crop 0.25 & Flip & Str 1.0 & Str 0.5 \\ 
\midrule
\multirow{4}{*}{Art Images} & \method{} (Trustmark + Dino) & 
\textbf{0.993} & \textbf{0.697} & \textbf{0.796} & \textbf{0.861} & 0.991 & \textbf{0.993} & \textbf{0.993} & 0.784 & \textbf{0.993} \\
& \method{} (Trustmark + CLIP) & \textbf{0.993} & 0.667 & 0.780 & 0.837 & \textbf{0.992} & \textbf{0.993} & \textbf{0.993} & 0.732 & \textbf{0.993} \\
& Naive IR (Dino) & 0.990 & 0.693 & 0.783 & 0.834 & 0.752 & 0.836 & 0.892 & \textbf{0.785} & 0.841 \\
& Naive IR (CLIP) &  0.990 & 0.659 & 0.759 & 0.795 & 0.771 & 0.806 & 0.856 & 0.732 & 0.801 \\
\midrule
\multirow{4}{*}{WikiArt} & \method{} (Trustmark + Dino) & \textbf{1.000} & 0.699 & \textbf{0.907} & \textbf{0.976} & \textbf{0.998} & \textbf{1.000} & \textbf{1.000} & \textbf{0.911} & 0.999  \\ 
& \method{} (Trustmark + CLIP) & \textbf{1.000} & \textbf{0.706} & 0.873 & 0.947 & \textbf{0.998} & \textbf{1.000} & \textbf{1.000} & 0.853 & \textbf{1.000}  \\ 
& Naive IR (Dino) & \textbf{1.000} & 0.698 & \textbf{0.907} & 0.971 & 0.974 & 0.985 & 0.995 & \textbf{0.911} & 0.985 \\ 
& Naive IR (CLIP) & \textbf{1.000} & \textbf{0.706} & 0.870 & 0.940 & 0.949 & 0.975 & 0.992 & 0.853 & 0.980 \\ 
\midrule
\multirow{4}{*}{ImageNet} & \method{} (Trustmark + Dino) & 
\textbf{1.000} & \textbf{0.875} & 0.975 & \textbf{0.992} & \textbf{1.000} & \textbf{1.000} & \textbf{1.000} & \textbf{0.946} & \textbf{1.000} \\
& \method{} (Trustmark + CLIP) & \textbf{1.000} & 0.698 & 0.841 & 0.934 & 0.995 & 0.999 & \textbf{1.000} & 0.840 & 0.999 \\
& Naive IR (Dino) & \textbf{1.000} & \textbf{0.875} & \textbf{0.976} & 0.991 & 0.983 & 0.997 & 0.999 & \textbf{0.946} & 0.996 \\
& Naive IR (CLIP) &  \textbf{1.000} & 0.697 & 0.836 & 0.926 & 0.894 & 0.929 & 0.993 & 0.840 & 0.980 \\
\midrule
& & Str 0.25 & Blur & Comb 0.5 & Comb 0.25 & Jitter & Diff 0.2 & Diff 0.15 & Diff 0.1 \\
\midrule
\multirow{4}{*}{Art Images} & \method{} (Trustmark + Dino) & \textbf{0.993} & \textbf{0.993} & \textbf{0.750} & \textbf{0.841} & \textbf{0.975} & \textbf{0.765} & \textbf{0.790} & \textbf{0.813} \\
& \method{} (Trustmark + CLIP) & \textbf{0.993} & \textbf{0.993} & 0.672 & 0.799 & 0.954 & 0.705 & 0.764 & 0.806 \\
& Naive IR (Dino) & 0.846 & 0.912 & 0.723 & 0.799 & 0.876 & \textbf{0.765} & \textbf{0.790} & \textbf{0.813} \\
& Naive IR (CLIP) & 0.818 & 0.847 & 0.638 & 0.746 & 0.770 & 0.705 & 0.764 & 0.806 \\
\midrule
\multirow{4}{*}{WikiArt} & \method{} (Trustmark + Dino) & \textbf{1.000} & \textbf{0.999} & \textbf{0.816} & \textbf{0.948} & \textbf{0.973} & \textbf{0.806} & \textbf{0.904} & \textbf{0.955} \\ 
& \method{} (Trustmark + CLIP) & \textbf{1.000} & \textbf{0.999} & 0.696 & 0.908 & 0.932 & 0.660 & 0.816 & 0.924 \\ 
& Naive IR (Dino) & 0.989 & 0.995 & 0.811 & 0.945 & 0.964 & \textbf{0.806} & \textbf{0.904} & \textbf{0.955} \\ 
& Naive IR (CLIP) & 0.988 & 0.977 & 0.692 & 0.905 & 0.862 & 0.660 & 0.816 & 0.924 \\ 
\midrule
\multirow{4}{*}{ImageNet} & \method{} (Trustmark + Dino) & \textbf{1.000} & \textbf{1.000} & \textbf{0.932} & \textbf{0.983} & \textbf{0.986} & \textbf{0.972} & \textbf{0.986} & \textbf{0.994} \\
& \method{} (Trustmark + CLIP) & \textbf{1.000} & 0.999 & 0.660 & 0.875 & 0.943 & 0.859 & 0.907 & 0.968 \\
& Naive IR (Dino) & \textbf{1.000} & 0.999 & 0.931 & \textbf{0.983} & 0.984 & \textbf{0.972} & \textbf{0.986} & \textbf{0.994} \\
& Naive IR (CLIP) & 0.992 & 0.974 & 0.648 & 0.858 & 0.852 & 0.859 & 0.907 & 0.968 \\
\bottomrule
\end{tabular}
\end{adjustbox}
\vspace{0.1cm}
\caption{Comparison between the robustness of \method{} and naive IR, with DinoV2 or CLIP embeddings. For a description of the modification parameters, refer to Appendix~\ref{appendix:augs}.}
\label{tab:accs}
\vspace{-0.4cm}
\end{table}

\section{Experiments}
\vspace{-0.2cm}
\label{sec:experiments}

In this section, we present experiments demonstrating \method{}'s effectiveness for source identification in the image domain, comparing its performance with the baseline of purely embedding-based image retrieval (which we call "naive IR").

\vspace{-0.1cm}
\subsection{Experimental Setup}

\textbf{Datasets.} Our experiments utilize a diverse range of image datasets, including ImageNet \citep{imagenet} (1.2M images), WikiArt \citep{wikiart} (80k images), PlantVillage \citep{plantvillage} (55k images), FGVC Aircraft \citep{fgvcaircraft} (10k images), Art Images \citep{artimages} (8k images), and Chest X-Ray Images \citep{chestxray} (5k images). For consistency, all images are resized and cropped to $256\times 256$ pixels.

\textbf{Embedding models.} We use DinoV2 \citep{dinov2} and CLIP/ViT-L-14 \citep{clip} to compute image embeddings for both our method and naive IR. Both models generate $768$-dimensional embeddings (with DinoV2's original size of $257\times768$ averaged over the first dimension).

\textbf{Watermarking models.} Our experiments primarily utilize Trustmark \citep{trustmark} due to its consistent robustness and minimal visible watermark patterns. We also report results using StegaStamp \citep{stegastamp} in the appendix (Figure~\ref{fig:acc_dinov2_stega}), noting that while Trustmark generally offers better robustness, StegaStamp excels against specific image modifications like diffusion purification \citep{saberi2024robustness}. We use $100$-bit watermarks (i.e., $n=100$), as higher bit counts can degrade image quality with current watermarking techniques.

\textbf{Error control codes (ECC).} For error correction, we employ Polar Codes \citep{polar_codes} with a successive cancellation decoder\footnote{\url{https://github.com/fr0mhell/python-polar-coding}}. The reliability of decoding is checked using the last log-likelihood ratio value, with a threshold set at $0.5$.

\textbf{Data modifications.} To evaluate robustness, we apply a range of common data augmentations, each parameterized to adjust its severity (e.g., crop.0.5 refers to center cropping images to $192\times192$ pixels and resizing back to $256\times256$). Detailed augmentation parameters are provided in Appendix~\ref{appendix:augs}. These augmentations ensure the images retain their semantic content.

\textbf{Evaluation.}  For evaluation, we sample $1000$ queries from each dataset and for each image modification. For an accurate evaluation, both our method and naive IR use exact nearest neighbor similarity search, as opposed to faster approximate nearest neighbor (ANN) techniques. We also evaluate out-of-dataset queries in Appendix~\ref{appendix:out-of-dataset} by sampling from the test sets of the datasets. We use $1024$ clusters in our experiments for \method{} ($k=10$).

\begin{figure}
    \centering
    \vspace{-0.1cm}
    \begin{minipage}[t]{.48\textwidth}
      \centering
      \includegraphics[width=1.\linewidth, trim={1cm 0.5cm 0.5cm 0}, clip]{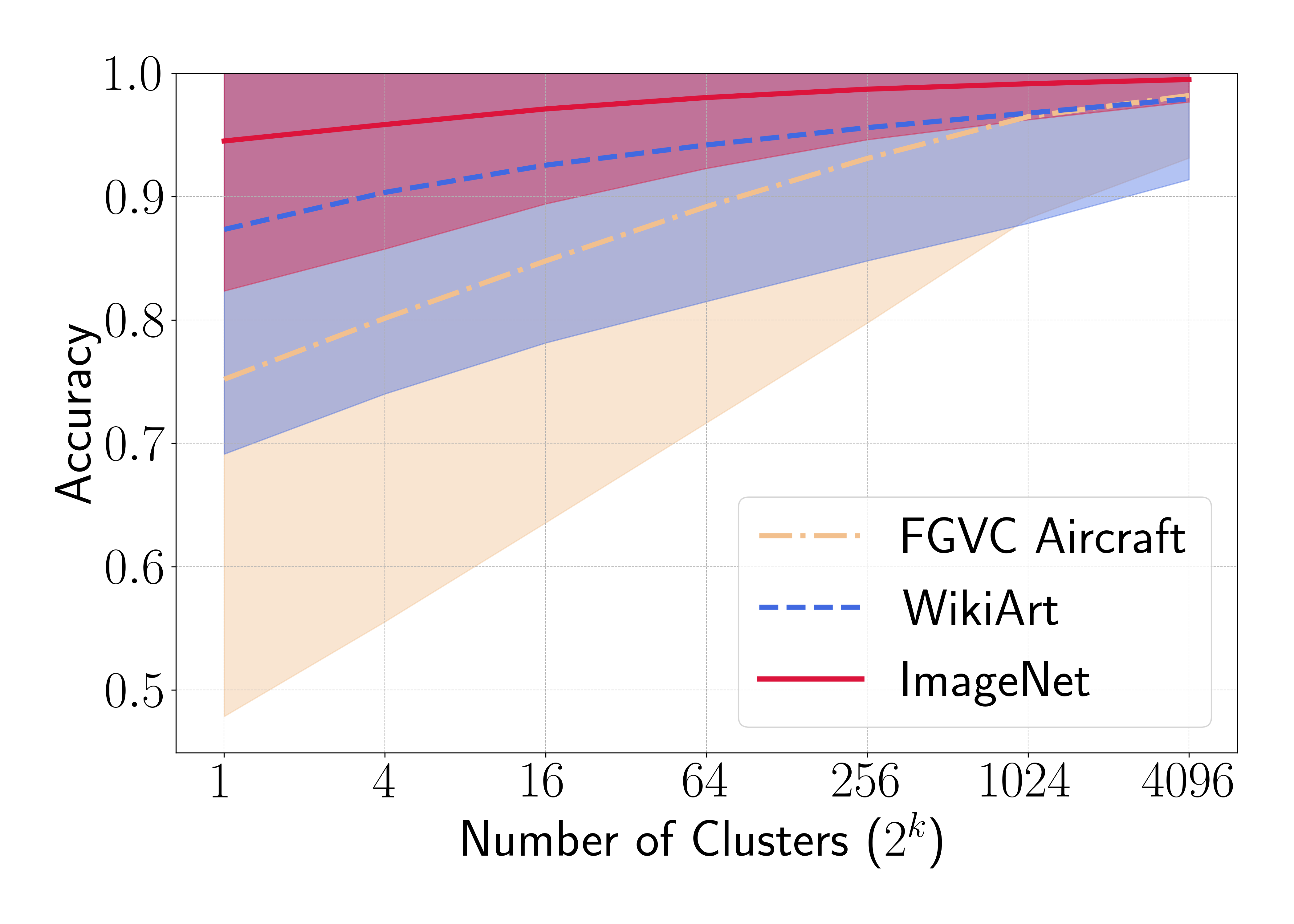}
      \vspace{-0.3cm}
      \captionof{figure}{Accuracy of image retrieval using DinoV2 embeddings on a subset of $\frac{N}{2^k}$ images from each dataset. The plot illustrates the mean (lines) and std (shades around the lines) of accuracy across a set of augmentations.
      }
      \label{fig:k_analysis}
    \end{minipage}%
    \hspace{0.2cm}
    \begin{minipage}[t]{.48\textwidth}
      \centering
      \includegraphics[width=1.\linewidth, trim={1cm 0.5cm 0.5cm 0}, clip]{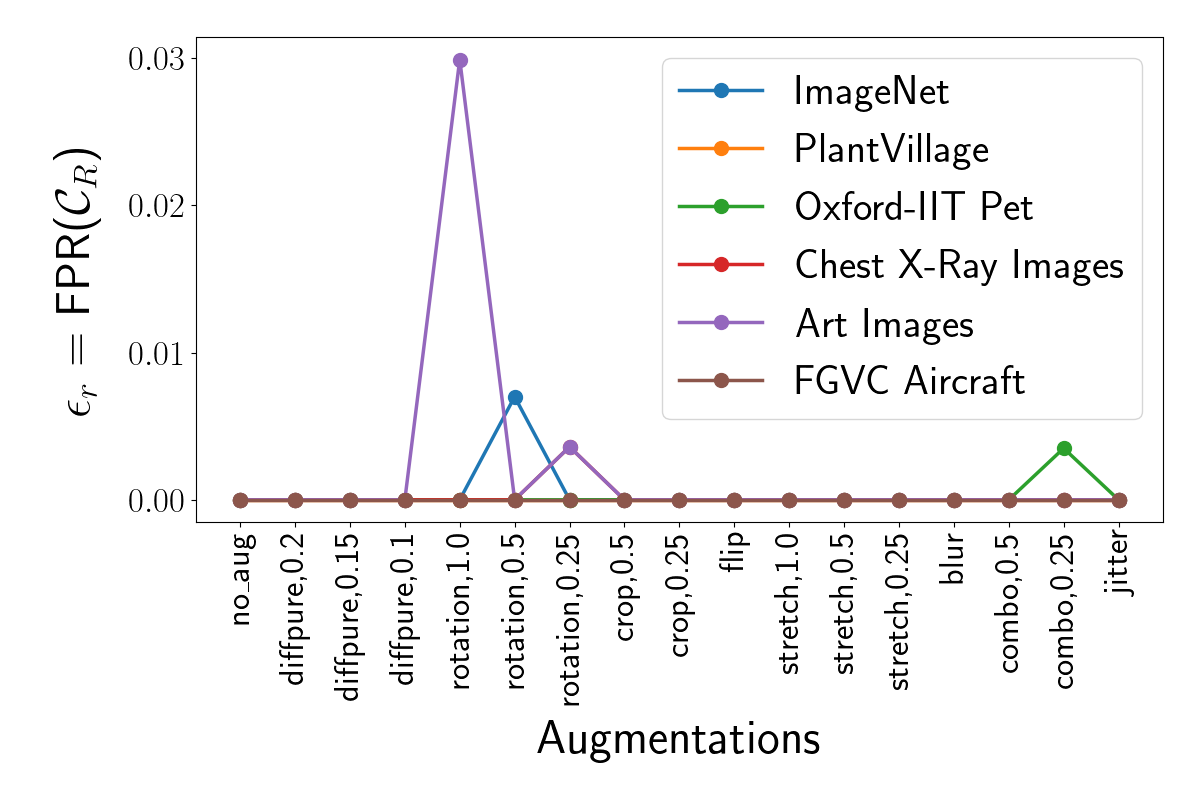}
      \vspace{-0.3cm}
      \captionof{figure}{False positive rate of the ECC's reliability check module ($\epsilon_r$), for various datasets and augmentations. This value was used in analysis in Section~\ref{sec:performance_analysis}.}
      \label{fig:epsilon_r}
    \end{minipage}
    \vspace{-0.3cm}
\end{figure}

\subsection{Results}

Figure~\ref{fig:acc_dinov2} compares the accuracy of \method{} (using Trustmark for watermarking) to naive IR in identifying matches with DinoV2 embeddings, on a set of image augmentations. The full set of augmentation results is shown in Figure~\ref{fig:acc_dinov2_full}, indicating no performance degradation even for less robust augmentations. Our results in Figure~\ref{fig:acc_dinov2} show that \method{}'s improvement over the baseline is less significant in datasets where DinoV2 is highly robust (e.g., ImageNet and WikiArt). This may be due to the more distinguishable nature of these images or the embedding models' prior exposure to similar data during training. However, for datasets like Art Images or Chest X-Ray Images, watermarking techniques generalize better to out-of-distribution or underrepresented data than embedding models, making \method{} more suitable for large-scale diverse data. Detailed results, including those for CLIP embeddings, are presented in Table~\ref{tab:accs} and visualized in Figure~\ref{fig:acc_clip}. Furthermore, Figure~\ref{fig:acc_clip} visualizes the results for CLIP embeddings.


Our experiments show that \method{} does not cause any noticeable performance degradation compared to the naive IR baseline. As depicted in Figure~\ref{fig:acc_dinov2}, Trustmark exhibits low robustness against "rotation" and "diffpure" modifications, likely due to the specific training regimen of the Trustmark model. In Figure~\ref{fig:acc_dinov2_stega}, we demonstrate that StegaStamp exhibits greater robustness to "diffpure" modifications, but performs weakly against other modifications.

A critical analysis raises whether \method{} can be enhanced with better watermarking techniques or if there is an inherent performance ceiling. Figure~\ref{fig:k_analysis} shows that with a perfectly robust watermarking method accurately identifying the ground truth cluster for query images, \method{}'s accuracy can improve significantly with practical $k$ values (without requiring excessive watermark bits). This suggests that $\mathbb{P}(x^=x_i , \bar{x}^ \neq x_i)$ is not the primary performance bottleneck in Equation~\ref{eq:improvement}, suggesting future research to focus on designing more robust watermarks to enhance \method{}'s overall robustness.

Furthermore, empirical results on the false positive rate of $\mathcal{C}_R$ are presented in Figure~\ref{fig:epsilon_r}. The false positive rate is minimal in most cases, justifying the practical applicability of \method{}, as this is the only scenario where \method{} might underperform compared to naive IR.

\section{Discussion and Limitations}

While our results demonstrate improved robustness compared to the baseline, we note that we are proposing a generalizable and scalable framework, which can be further enhanced by the rapid ongoing advancements in watermarking techniques and data embedding models. To highlight the rapid progress in watermarking techniques, the Trustmark \citep{trustmark} method that we utilize in our framework, is a recently proposed method that has shown significant overall robustness compared to the previous state-of-the-art image watermarks.

As shown in the literature \citep{saberi2024robustness, rabbani2024waves}, watermarks with a higher perturbation level (i.e., more visible watermark traces) can substantially improve robustness against non-adversarial attacks (e.g., common augmentations, diffusion purification). While our framework may have limited robustness when using imperceptible watermarking techniques, in some scenarios, higher perturbation watermarks can be utilized for enhanced protection. For example, to protect the ownership of images posted on social media platforms, users can choose to have more visible watermark traces in their media, benefiting from greater content protection. 

In our experiments, we evaluate robustness against simple image modifications. However, more complex modifications that do not change the semantics of the image can also be applicable (e.g., text insertion, inpainting). For some modifications, our framework can potentially provide more substantial robustness enhancements. For instance, consider inserting an image from the dataset into a larger random image. While naive retrieval may struggle to find a match for this type of modification, our framework can use watermark detectors \citep{stegastamp} to first detect and extract the region of the query image containing the watermark, and then pass the extracted region to the retrieval module. We leave the analysis of other modification types to future work.







\section{Conclusion}

In this paper, we introduced \method{} (\methodexpand{}), a novel framework that combines error-controlled watermarking with embedding-based retrieval for source identification in data provenance.
This novel approach addresses the limitations inherent in current methods, such as the restricted capacity of watermarking and the scalability issues of embedding-based retrieval in large databases. Furthermore, by employing ECC algorithms, our method ensures robust and reliable utilization of watermarks, even under extensive data modifications.
In our experiments, \method{} demonstrates improved robustness compared to traditional retrieval, particularly in challenging datasets. This makes our framework a reliable and effective alternative for source identification.

\section{Acknowledgement}
This project was supported in part by a grant from an NSF CAREER AWARD 1942230, ONR YIP award N00014-22-1-2271, ARO’s Early Career Program Award 310902-00001, HR00112090132 (DARPA/RED), HR001119S0026 (DARPA/GARD), Army Grant No. W911NF2120076, the NSF award CCF2212458, NSF Award No. 2229885 (NSF Institute for Trustworthy AI in Law and Society, TRAILS), an Amazon Research Award and an award from Capital One.

\bibliography{refs}
\bibliographystyle{refs}

\newpage

\appendix


\begin{figure}
    \centering
    \includegraphics[width=1.0\textwidth]{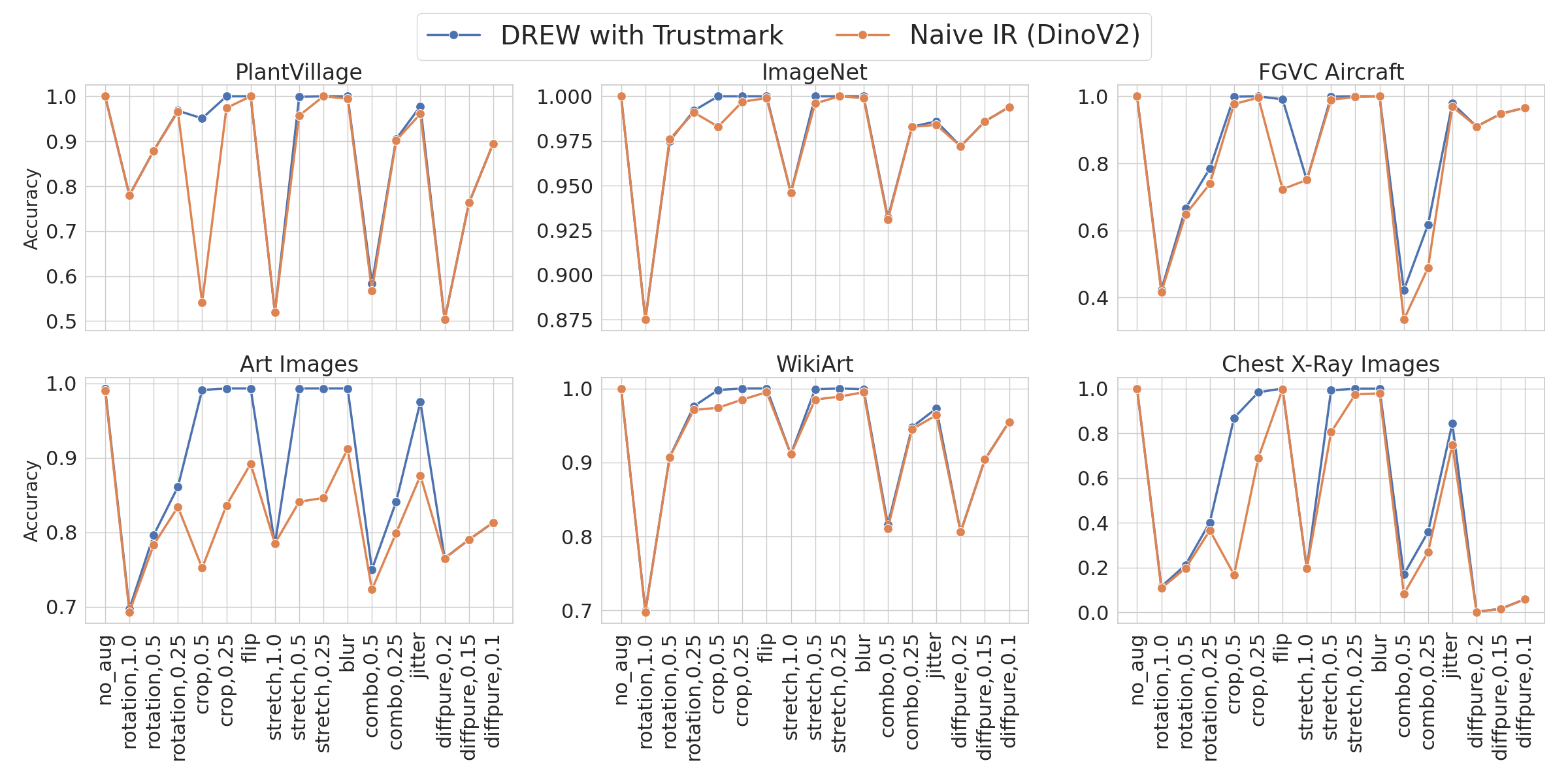} \\
    \caption{Accuracy of our method vs naive retrieval using DinoV2 embeddings, against different types and severity levels of augmentations. The utilized watermarking method is Trustmark.}
    \label{fig:acc_dinov2_full}
    \vspace{-0.3cm}
\end{figure}

\begin{figure}
    \centering
    \includegraphics[width=1.0\textwidth]{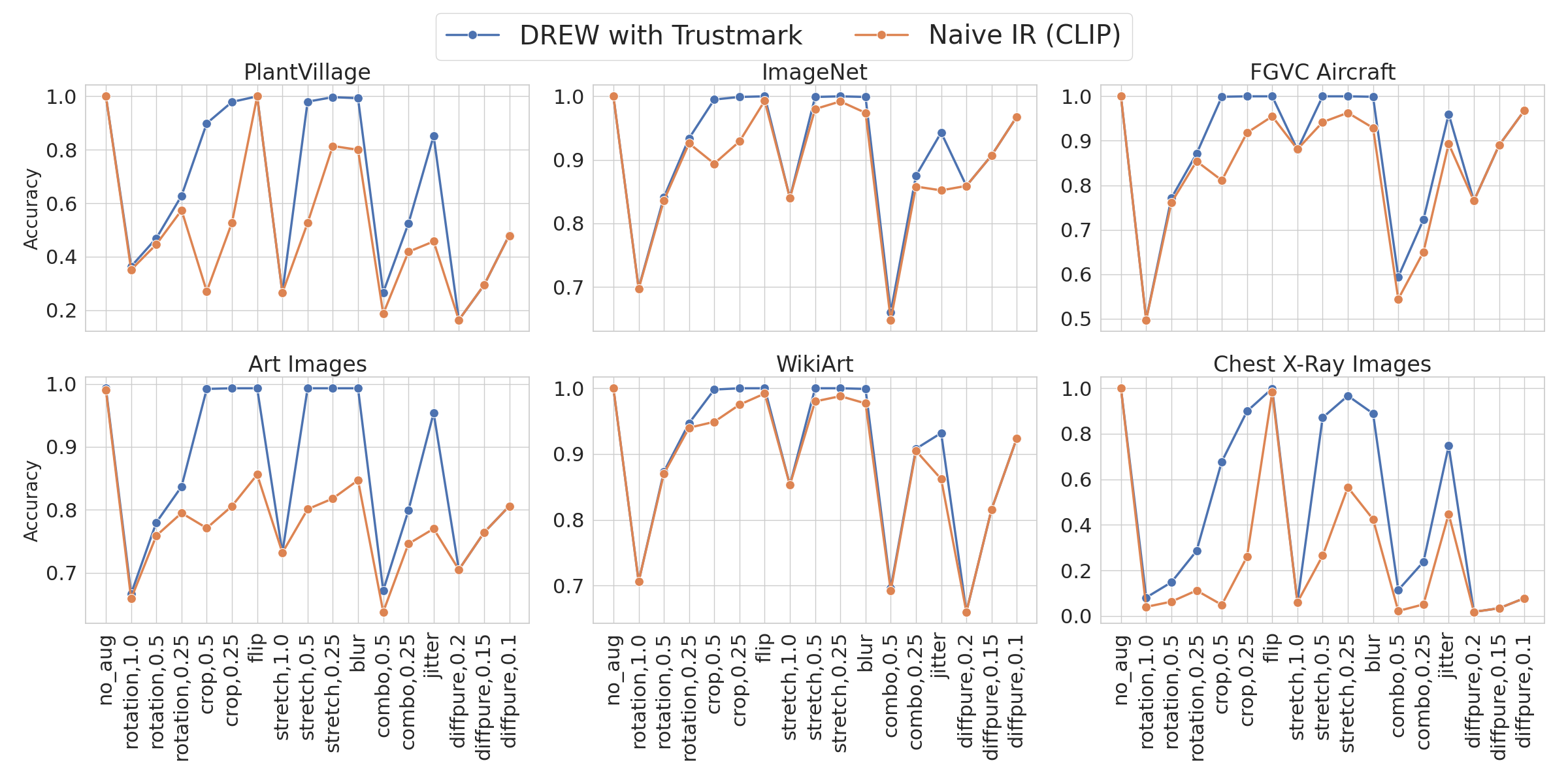} \\
    \caption{Accuracy of our method vs naive retrieval using CLIP/ViT-L-14 embeddings, against different types and severity levels of augmentations. The utilized watermarking method is Trustmark. In general, CLIP shows lower robustness than DinoV2 (see Figure~\ref{fig:acc_dinov2}), and our method provides a higher accuracy boost to the naive retrieval using CLIP.}
    \label{fig:acc_clip}
\end{figure}

\begin{figure}
    \centering
    \includegraphics[width=1.0\textwidth]{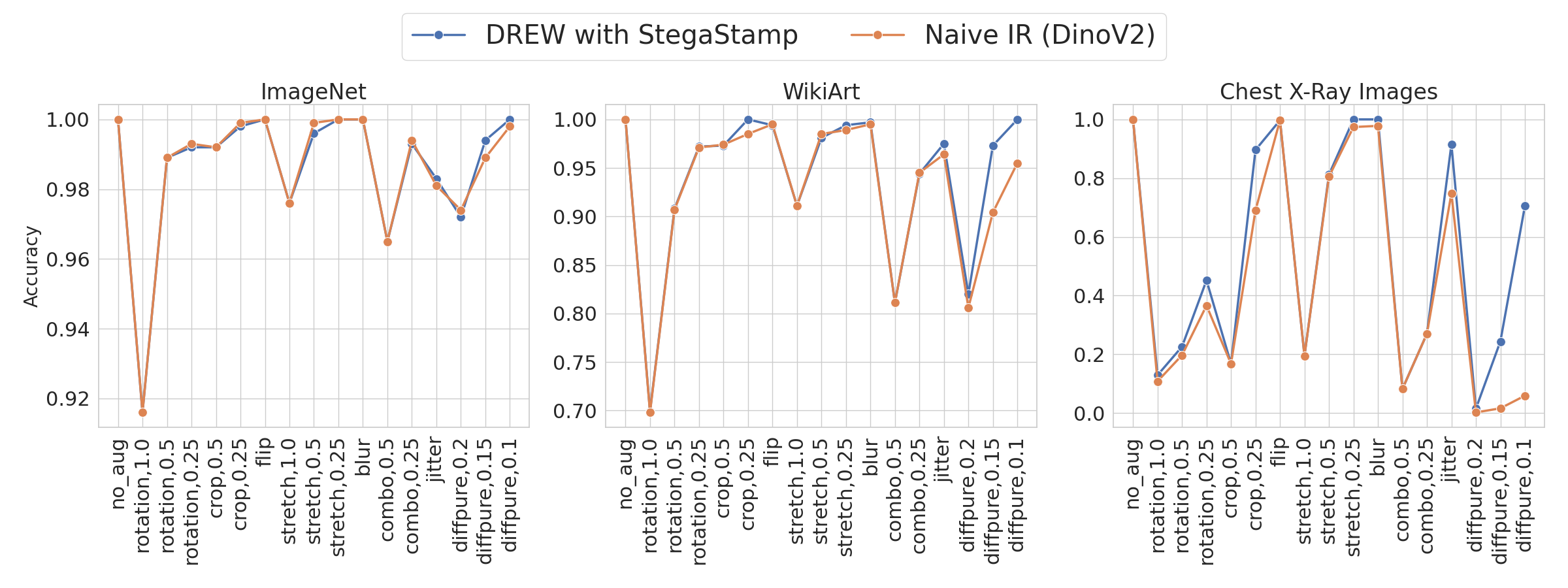} \\
    \caption{Accuracy of our method, implemented with StegaStamp watermark, vs naive retrieval using DinoV2 embeddings, against different types and severity levels of augmentations. StegaStamp performs better than Trustmark (see Figure~\ref{fig:acc_clip}) on "diffpure", while showing lower robustness to other image modifications.}
    \label{fig:acc_dinov2_stega}
\end{figure}







\begin{figure}
    \centering
    \includegraphics[width=1.0\textwidth]{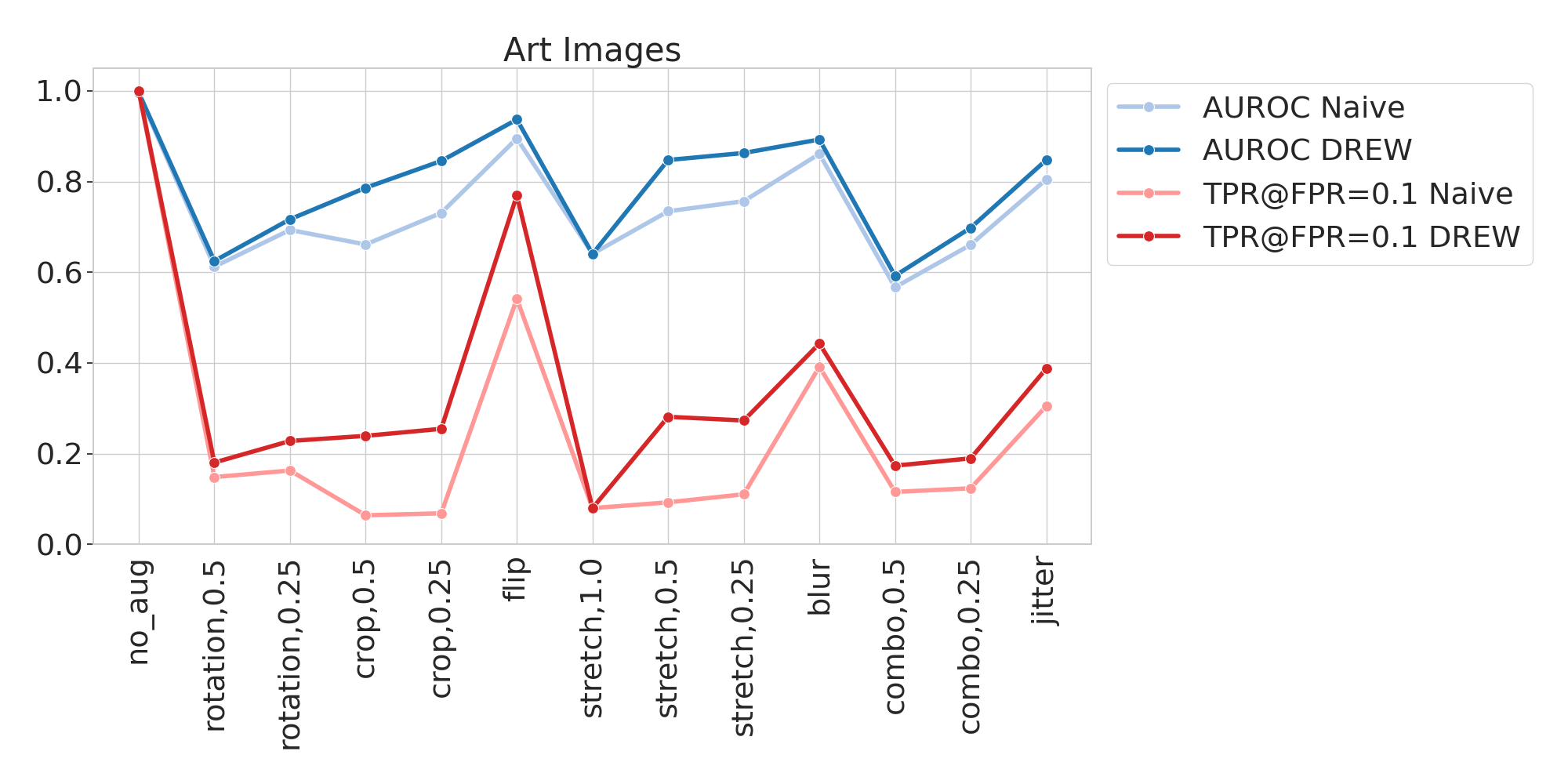} \\
    \caption{AUROC and TPR@FPR=$0.1$ (i.e., True positive rate at $0.1$ false positive rate) for \method{} vs naive IR. DinoV2 embeddings and Trustmark watermark are used for these results.}
    \label{fig:roc_dino}
\end{figure}

\section{Performance on out-of-dataset queries}
\label{appendix:out-of-dataset}

In section~\ref{sec:prelims}, we assumed that a retrieval method $\mathcal{M}$, reports a match for every query. But in some applications, the method should prevent reporting false positive matches. For \method{}, similar to the traditional data retrieval, we consider the embedding similarity $\phi(x_i)^T \phi(z)$ between the query $z$, and the match $x_i$, and define a threshold $t_{sim}$ to distinguish between true matches (i.e., $\phi(x_i)^T \phi(z) \geq t_{sim}$) and false matches (i.e., $\phi(x_i)^T \phi(z) < t_{sim}$).

For out-of-distribution queries that are not supposed to match any data from the dataset, if \method{} cannot find a reliable cluster for the query, it reports the same match as naive IR. Otherwise, if \method{} falsely assigns a cluster to the query, it will report the data sample with the highest embedding similarity in that cluster, whose similarity score is less than the reported output of naive IR (i.e., $max_{x \in X_c} \phi(x)^\top \phi(z) \leq max_{x \in X} \phi(x)^\top \phi(z)$). Therefore, even though the second case does not often happen in practice (due to the robustness of ECC), in cases where it happens, it will even further boost the performance of \method{}.

In Figure~\ref{fig:roc_dino}, we evaluate \method{} on a combination of $1000$ in-dataset and $1000$ out-of-dataset queries, and compare the results to naive IR.

\section{ECC Analysis}
\label{appendix:ecc}

Assume $x \in X$ and $z=\mathcal{A}(x)$ is the altered version of $x$. Passing these data samples into the watermark decoder results in $n$-bit binary keys $w_x=\mathcal{W}_D(x)$ and $w_z=\mathcal{W}_D(z)$. Based on the robustness of the watermark, $w_x$ and $w_z$ can have different bits in several places. For the current analysis, we assume that an attack $\mathcal{A}$ causes each bit of $w_z$ to flip independently from their value in $w_x$, with probability $p_{\mathcal{A}}$. Note that in practice, this assumption does not necessarily hold for every attack and watermarking technique.

In Section~\ref{sec:performance_analysis}, we have provided discussions on the correlation between outputs of $\mathcal{C}_D$ and $\mathcal{C}_R$, and their effect on the performance of the model. In this section, we analyze the robustness of $\mathcal{C}_D$ against attacks. In other words, if $c_x=\mathcal{C}_D(w_x)$ and $c_z=\mathcal{C}_D(w_z)$, we are interested in the success rate of $\mathcal{C}_D$, which is the cases where $c_z=c_x$.

As defined in Section~\ref{sec:prelims}, $\mathcal{C}_D$ receives a $n$-bit watermark key, and outputs a $k$-bit cluster code where $k < n$. The \textit{code rate} of the ECC module $\mathcal{C}$ is defined as $\frac{k}{n}$. Generally, a lower code rate corresponds to more redundant bits used for correction, and will result in a higher decoding success rate. Shannon's capacity theorem \citep{shannon_theorem}, provides an upper-bound on the code rate of an ECC algorithm with a high probability of successful decoding against an attacker that flips the bits of the $n$-bit keys with probability $p_{\mathcal{A}}$:

\begin{equation}
\label{eq:shannon_limit}
    \frac{k}{n} \leq 1-H(p_{\mathcal{A}}) =  1+p_{\mathcal{A}} log(p_{\mathcal{A}})+(1-p_{\mathcal{A}}) log(1-p_{\mathcal{A}}),
\end{equation}

where $H$ is the binary entropy function.

\begin{figure}
    \centering
    \includegraphics[width=0.8\textwidth]{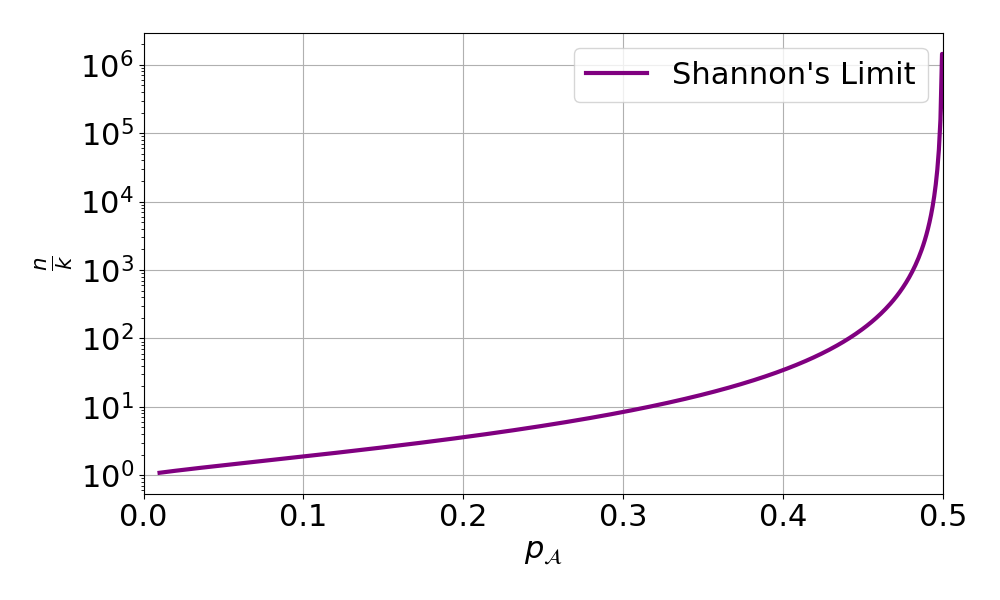} \\
    \caption{Illustration of Equation~\ref{eq:shannon_limit}, which shows the amount of bit redundancy (i.e., $\frac{n}{k}$) required for a successful decoding against $p_\mathcal{A}$ bit-flip rate w.h.p. In our experiments in Section~\ref{sec:experiments}, we use $k=10, n=100$.}
    \label{fig:shannon_limit}
\end{figure}

In practice, state-of-the-art ECC algorithms such as polar codes \citep{polar_codes} and LDPC \citep{ldpc}, achieve code rates that are close to the Shannon's capacity limit. Figure~\ref{fig:shannon_limit} illustrates the limit from Equation~\ref{eq:shannon_limit}. In practice, there is a limit to the number of bits that can be encoded into a data sample (e.g., image) as a watermark. Therefore, to increase the robustness of the $\mathcal{C}_D$, the solution is to decrease $k$ , which corresponds to decreasing the number of clusters for our method. This introduces a trade-off between the terms $P(c=c_{gt})$ and $P(x^*=x_i , \bar{x}^* \neq x_i)$ that have been used in the analysis from Section~\ref{sec:performance_analysis}.

\section{Robustness to Adversarial Attacks}
\label{appendix:adversarial}

In this section, we discuss the performance of \method{} in various adversarial scenarios. Since the data retrieval methods we discuss in this work rely on using embeddings from neural networks, they might be susceptible to adversarial attacks. We discuss three possible scenarios below:

\begin{enumerate}
\item{
Let $x_i$ be an image in the dataset $X$. The adversary could imperceptibly perturb $x_i$ as $\tilde{x}_i = x_i+\delta$ where $\delta = \min_{\|\delta\| \leq \epsilon} \phi(x_i)^\top \phi(\tilde{x}_i)$. In this case, the retrieval technique based on embedding similarity match would return another image $x_j$ as a match for the query $\tilde{x}_i$ instead of the datapoint $x_i$, where $\phi(x_j)^\top \phi(\tilde{x}_i) > \phi(x_i)^\top \phi(\tilde{x}_i)$. However, this attack scenario could affect both \method{} and the baseline embedding-based retrieval techniques to the same extent.
}
\item{
In another scenario, the adversary could add imperceptible noise $\delta = \min_{\|\delta\| \leq \epsilon} \mathcal{C}_R(\mathcal{W}_D(x_i+\delta))$ such that the output of the error control decoder is flagged as unreliable. For this case, the ECC-based watermark signal is removed from the query image $\tilde{x}_i = x_i + \delta$, and hence, would lead to \method{} searching the entire database for a match. Note that this attack only makes \method{} as effective as the baseline embedding-based retrieval techniques and not any worse.
}
\item{
The adversary could \textit{spoof} the datapoint $x_i$ in cluster $X_i$ by adding adversarial noise such that, while querying the perturbed $\tilde{x}_i = x_i + \delta$, \method{} would search for the query candidate $\tilde{x}_i$ in a different cluster $X_j$. Here $\delta$ should satisfy $\min_{\|\delta\|\leq \epsilon, j \neq i} |\mathcal{C}_D(\mathcal{W}_D(x_i + \delta)) - j| - \mathcal{C}_R(\mathcal{W}_D(x_i + \delta))$. In this case, \method{} would confidently search for $\tilde{x}_i$ in a cluster $X_j$ and fail to return the matching datapoint $x_i$ present in cluster $X_i$.  
}
\end{enumerate}

Note that the attack scenarios mentioned here could also be practical in a black-box fashion. 
In Scenario 1, the adversary could use a surrogate embedding model instead of $\phi$ to perform their optimization to attack the embedding-based matching.
For scenarios 2 and 3, as shown in \citep{saberi2024robustness}, the adversary could use a dataset of watermarked samples from the dataset $X$ and non-watermarked samples not present in the dataset $X$ to learn a classifier. 
The adversary can then attack the datapoints to erase or spoof watermarks for a query datapoint based on this classifier's confidence. As we discuss here, Scenarios 1 and 2 affect \method{} and the baseline methods to the same level. 
Scenario 3 could lead \method{} to yield a degraded performance.
However, for this setting, the adversary should assume having access to sufficient data points from at least two different clusters, source $X_i$ and target clusters $X_j$.
This may not be practical in all scenarios.
For this reason, in this application, spoofing is not interesting for a practical adversary since they would resort to using a surrogate embedder, as mentioned in Scenario 1, to adversarially inhibit accurate retrieval.
We leave studying \method{} to be robust in Scenario 3 for the future.

\section{Data Modifications}
\label{appendix:augs}

In this section, we provide a code for a detailed description of the used image augmentations and their parameters.

\begin{verbatim}
# inputs: 
# aug
# severity (in range [0, 1])
if aug == 'no_aug':
    transform = transforms.Compose([])
elif aug == 'rotation':
    transform = transforms.Compose([
            transforms.RandomRotation([-int(90 * severity), int(90 * severity)]),
    ])
elif aug == 'crop':
    transform = transforms.Compose([
            transforms.CenterCrop(256 - int(128 * severity)),
            transforms.Resize((256, 256)),
    ])
elif aug == 'flip':
    transform = transforms.Compose([
            transforms.RandomHorizontalFlip(1.0),
    ])
elif aug == 'stretch':
    transform = transforms.Compose([
            transforms.Resize(256),
            transforms.CenterCrop((256 - int(128 * severity), 256)),
            transforms.Resize((256, 256)),
    ])
elif aug == 'blur':
    transform = transforms.Compose([
            transforms.GaussianBlur(5),
            # transforms.ToTensor(),
    ])
elif aug == 'combo':
    transform = transforms.Compose([
            transforms.CenterCrop((256 - int(128 * severity), 256)),
            transforms.RandomHorizontalFlip(1.0),
            transforms.RandomRotation([-int(90 * severity), int(90 * severity)]),
            transforms.Resize((256, 256)),
    ])
elif aug == "jitter":
    transform = transforms.Compose([
            transforms.ColorJitter(brightness=1.0, contrast=0.5, saturation=1.0, hue=0.1),
    ])
elif aug == 'diffpure':
    transform = transforms.Compose([
        DiffPure(steps=severity),
    ])
\end{verbatim}

\end{document}